\documentstyle[twoside,fleqn,espcrc2]{article}

\input epsf
\def\Re{\mbox{Re}}
\def\Im{\mbox{Im}}

\newcommand{\AmS}{{\protect\the\textfont2
  A\kern-.1667em\lower.5ex\hbox{M}\kern-.125emS}}

\title{Weak matrix elements: On the way to $\Delta I=1/2$ rule and
$\varepsilon '/\varepsilon$ with staggered fermions.}

\author{D. Pekurovsky and G. Kilcup\address{Department of Physics,
the Ohio State University, \\ 174 W. 18th Ave., Columbus OH 43210, USA}}
       
\begin{document}

\begin{abstract}
We report progress in our study of hadronic weak matrix elements 
relevant for the $\Delta I=1/2$ rule and $\varepsilon '/\varepsilon$.
The presented results are from our first runs on 
a quenched ensemble with \mbox{$\beta =6.0$}, and a dynamical 
\mbox{$N_f=2$} ensemble with \mbox{$\beta =5.7$,} using staggered 
gauge-invariant tadpole-improved fermionic operators. 
\end{abstract}

\maketitle

\section{Introduction}
An important contribution of Lattice QCD to phenomenology is
a first-principle calculation of non-perturbative matrix
elements (ME's) of weak operators involving light hadrons. Knowledge
of these ME's combined with experimental data translates directly to
constraints on CKM matrix elements, and thus enables another test
of the Minimal Standard Model.

In this work we concentrate on weak decays of kaons into two pions. 
$\varepsilon '$ is the measure of direct CP violation in these 
decays. It is defined as $\varepsilon ' = i\Im A_2 e^{i(\delta_2 - 
\delta_0)}/\sqrt{2}A_0$, where $A_2$, $\delta_2$, $A_0$ and $\delta_0$
are amplitudes and final interaction phases corresponding to isospin
2 and 0 final pion state.  
New experiments will soon measure $\varepsilon '$,
hopefully resolving the currently muddy situation.
In addition to calculating $\varepsilon '$, it is interesting whether
Lattice QCD can 
adequately explain the dominance of $\Delta I=1/2$ transitions
in kaon decays, i.e. the fact that 
$\omega\equiv \Re (A_0/A_2) = 22.2$.

\section{Method}
We work within the framework of an effective field theory obtained
by integrating out W-boson and t-quark, using OPE and running down
the effective Hamiltonian to scales of order of the lattice scale using 
RG equations \cite{buras}. At these scales the effective Hamiltonian
has the form
\begin{equation}
H_W^{\it eff} = 
\frac{G_F}{\sqrt{2}} V_{ud}\,V^*_{us} \sum_{i=1}^{10} \Bigl[
z_i(\mu) + \tau y_i(\mu) \Bigr] O_i (\mu) 
 \, , 
\end{equation}
where $\tau = - V_{td}V_{ts}^{*}/V_{ud} V_{us}^{*}$, $z_i$ and $y_i$
are Wilson coefficients (currently known at two-loop order), 
and $O_i$ is the basis of 10 four-fermions operators.
In this effective theory $\varepsilon '$ can be expressed in terms
of CKM matrix elements and $\langle \pi\pi | O_i | K\rangle $. 
Calculating the long-distance part of these ME's is the task for lattice
theorists. 

It happens that the value of $\varepsilon '$ is dominated by competing
contributions of the following two operators: 
\begin{eqnarray}
O_6 & = & (\bar{s}_\alpha d_\beta )_{V-A}\sum_q(\bar{q}_\beta q_\alpha )_{V+A}  \\
O_8 & = & \frac{3}{2}(\bar{s}_\alpha d_\beta )_{V-A}\sum_q e_q (\bar{q}_\beta q_\alpha )_{V+A} 
\end{eqnarray}
In this talk we concentrate mostly on calculation of ME's involving $O_6$. 

In general, we follow the technique of calculating ME's with staggered 
fermions developed in Ref.~\cite{kilcup1}. This method has proved 
to be successful in the past in calculation of the $B_K$ parameter, 
which enters the expression for indirect CP-violation parameter $\epsilon$. 
Shown in Fig.~1 are our latest results. During the last year, a new
point at $\beta =6.4$ has been added. The lattice spacings are
determined by demanding asymptotic scaling, with the overall scale
taken from the continuum limit of the $\rho$ mass.
The final result is $B_K(NDR,2 {\rm GeV})=0.552 \pm 0.007$.

\begin{figure}[htb]
\begin{center}
\leavevmode
\vspace{-1cm}
\centerline{\epsfysize=5cm \epsfbox{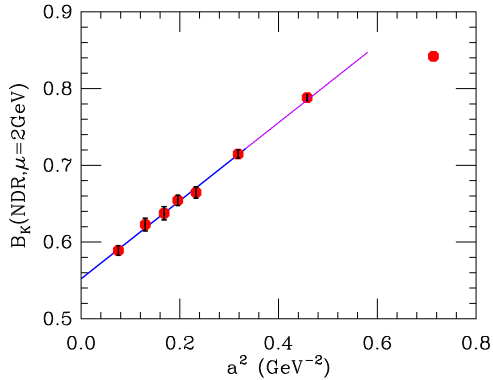}}
\vspace{-1.5cm}
\end{center}
\label{Bk}
\caption{Quenched $B_K$ using tadpole improved, gauge-invariant 
operators.
}
\vspace{-0.9cm}
\end{figure} 

We would like
to achieve similar precision and finesse with much noisier ME's
relevant for $\Delta I=1/2$ rule and $\varepsilon '$. We introduce
a number of improvements compared to the original work on
these ME's~\cite{kilcup2}.

Due to technical complications, it is extremely difficult to extract 
four-point functions on the lattice.
Instead, we calculate $\langle \pi|H_W|K\rangle$ and $\langle 0|H_W|K\rangle$
and (following Ref.~\cite{bernard}) use chiral perturbation theory to 
relate them to $\langle \pi\pi|H_W|K\rangle$.

It is convenient to calculate B-ratios of ME's, which are defined as ratios
of ME's to their values obtained by vacuum saturation.
We have to consider three types of 4-fermion contractions:
`figure-eight', `eye' and `annihilation'. The latter two are notoriously
noisy. See Ref.~\cite{kilcup1} for more details.

\section{Simulation}

\begin{table*}[hbt]
\setlength{\tabcolsep}{1.5pc}
\newlength{\digitwidth} \settowidth{\digitwidth}{\rm 0}
\catcode`?=\active \def?{\kern\digitwidth}
\caption{Simulation parameters}
\label{params}
\begin{tabular*}{\textwidth}{lllllll}
\hline
$\beta $ & $N_f$ &\# conf.&Lattice size& $a^{-1}$,GeV & Generated by\\
\hline
6.0 & 0 & 216 & $16^3\times (32\times 4)$ & 2.1 &OSC \\
5.7 & 2 & 71 & $16^3\times (32\times 4)$ & 2.0 &Columbia \\
\hline
\end{tabular*}
\end{table*}

Our simulation parameters are shown in Table~\ref{params}. 
We use periodic boundary conditions, and
replicate the lattice by a factor of 4 in the time direction. 
Only degenerate mesons ($m_s=m_d=m_u$) are considered. 
We employ gauge-invariant, tadpole-improved
staggered fermion operators.

\begin{figure}[htb]
\begin{center}
\leavevmode
\vspace{-1cm}
\centerline{\epsfysize=5cm \epsfbox{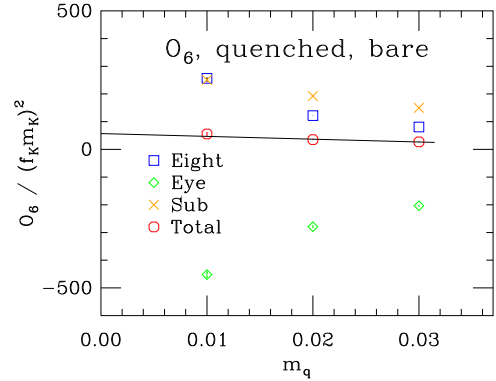}}
\vspace{-1.5cm}
\end{center}
\label{chiral}
\caption{Chiral plot of ME involving $O_6$ operator for quenched ensemble, 
without perturbative corrections. Three contributions come from 
`figure-eight', `eye' and `annihilation' contractions.} 
\vspace{-0.5cm}
\end{figure} 

In Fig.~2 we show three contributions to the (bare) ME of $O_6$ on
a chiral plot. 
Although individual contributions to $O_6$ might be diverging
in chiral limit, their sum total seems to converge to a finite value
with linear dependence on $m_q$, in agreement with 
chiral symmetries of staggered fermions.

\section{Matching with continuum}

To quote continuum results we use the `horizontal' matching procedure:
using $g_{\overline{MS}}$ as our
renormalized coupling constant, we perform the matching to continuum
at a scale $q^*$ with one-loop perturbation theory \cite{sharpe,ishizuka}:
\begin{eqnarray}
\displaystyle
O_i^{\it cont}(q^*) = & O_i^{\it lat} + \displaystyle\frac{g^2(q^*a)}{16\pi^2}\displaystyle\sum_j(\gamma_{ij}\ln (\frac{q^*a}{\pi} ) \nonumber \\
& + C_{ij})O_j^{\it lat} + O(g^4) + O(a^n) 
\end{eqnarray}
and then run to a specified final scale \mbox{$\mu$ 
(e.g. 2 GeV)} using continuum two-loop equations. We work in NDR 
variant of $\overline{\mbox{MS}}$ scheme.

The finite coefficients $C_{ij}$ are uncomfortably big for some operators.
In fact, for operators of type $PP$ the perturbative correction
is almost the size of the tree-level value.  This makes one-loop
perturbation theory unreliable. Directly related to this is a huge
dependence on $q^*$, which serves as an estimate of the size of
second-order corrections. 

The situation can be made slightly better if we follow the
continuum convention and quote results in terms of a $B$ parameter.
While there are still potentially large perturbative corrections,
we can minimize their effect by considering the $B$ parameter for
the operator $O_6/Z_P^2$.  This choice modifies
both the matching and anomalous dimension matrices, removing
the largest corrections from the dominant $PP$ matrix elements,
at the expense of larger corrections in front of subdominant operators.
As shown in figure 3, the remaining $q^*$ dependence is then modest.

\begin{figure}[htb]
\begin{center}
\leavevmode
\vspace{-1cm}
\centerline{\epsfysize=5cm \epsfbox{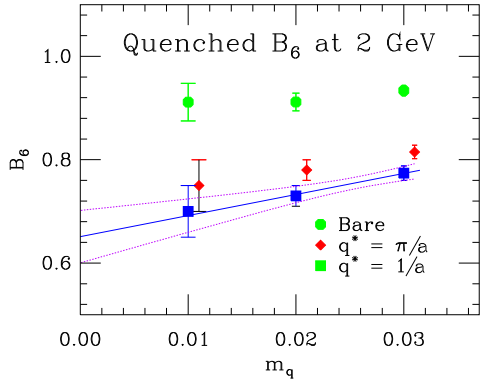}}
\vspace{-1.5cm}
\end{center}
\label{qstar}
\caption{$B_6$ ratio for quenched ensemble. The three groups of points
correspond to bare operators, and those renormalized in NDR scheme at 2 GeV
using two different values for $q^*$.} 
\vspace{-0.5cm}
\end{figure} 
Of course, this sleight of hand doesn't solve the remaining problem,
which is the apparent breakdown of perturbation theory for the
pseudoscalar renormalization $Z_P$, or equivalently, the
quark mass renormalization.
Once this coefficient is found (e.g. non-perturbatively) the value
of $O_6$ will be readily available. The situation
with other operators is the same. 

For the operators $SS_1$, $SS_2$ and $PP_1$ the renormalization
coefficients are not yet known.  For the central values
we have therefore taken the coefficients appropriate for
Landau gauge operators.  Varying their values by 100\%
we find that the change in $B_6$ and $B_8$ is insignificant.
For other operators (namely $B_5$ and $B_7$) 
we will eventually need to determine the missing
perturbative coefficients.

\section{Results and conclusions}

Our preliminary results for $B_6(\mbox{NDR, 2 GeV})$ at given values
of $\beta$ are 
$0.67 \pm 0.04 \pm 0.05$ for quenched
and $0.76\pm 0.03 \pm 0.05$ for dynamical ensembles (first error is
statistical, the second one is an estimate of higher-order perturbative 
corrections). For $B_8^{3/2}(\mbox{NDR, 2 GeV})$ we find 
$1.082 \pm 0.006 \pm 0.011$ for
quenched and $1.12 \pm 0.02 \pm 0.01$ for dynamical ensembles,
in reasonable agreement with results using Landau gauge and smeared
operators \cite{KGS}.

We do see a dominance of $\Delta I=1/2$ transition over $\Delta I=3/2$ one.
The ratio of amplitudes varies sensitively with the quark mass,
but at $ma=.01$, our preliminary results are:
$\omega\equiv \Re A_0/A_2 = 11 \pm 1 \pm 3$ for quenched
and $\omega = 16 \pm 3 \pm 3$ for dynamical ensemble. 

We have made a first step in a program to calculate $\varepsilon '$
and $\Delta I=1/2$ rule on the lattice. Apart from the common Lattice QCD
problems of (partial) quenching, finite lattice spacing and size
and degenerate quark masses, we have to face two more problems:
failure of perturbation theory, and mostly unknown uncertainty
in chiral perturbation theory predictions. To solve the first problem,
a non-perturbative determination of $Z_P$ is needed. 
We plan to repeat the calculations for a number of different $\beta$'s
in order to take the continuum limit.

We thank the Columbia group for access to the dynamical
gauge configurations, and the Ohio Supercomputer Center for the 
necessary Cray-T3D time.

\end{document}